\def\be{\begin{equation}}
\def\te{\end{equation}}
\def\bea{\begin{eqnarray}}
\def\nn{\nonumber}
\def\tea{\end{eqnarray}}

\def\a{\alpha}
\def\b{\beta}

\def\e{\epsilon}

\def\l{\lambda}

\def\o{\omega}

\def\r{\rho}

\def\t{\tau}

\def\O{\Omega}

\newskip\humongous \humongous=0pt plus 1000pt minus 1000pt
\def\caja{\mathsurround=0pt}
\def\eqalign#1{\,\vcenter{\openup2\jot \caja
        \ialign{\strut \hfil$\displaystyle{##}$&$
        \displaystyle{{}##}$\hfil\crcr#1\crcr}}\,}
\newif\ifdtup

\def\ha{{1\over 2}}

\documentstyle[12pt]{article}

\makeatletter                    
\@addtoreset{equation}{section}  
\makeatother                     

\begin{document}

\title{Quantum Noise in Gravitation and Cosmology}

\author{B. L. Hu\thanks{ Email: hu@umdhep.umd.edu}\\
{\small Department of Physics, University of Maryland,
College Park, MD 20742, USA} \\ A. Matacz\thanks{ Email:
amatacz@physics.adelaide.edu.au}\\
{\small Department of Physics, University of Adelaide, 5005, Australia}}
\maketitle
\centerline{\it Invited talk delivered by B. L. Hu at the Workshop on
Noise and Order,}
\centerline{\it Los Alamos National Laboratory, September, 1993}
\centerline{(umdpp 94-44) }

\begin{abstract}
We begin by enumerating the many processes in gravitation and cosmology where
quantum noise and fluctuations play an active role
such as particle creation, galaxy formation and entropy generation.
Using the influence functional we first explain the origin and
nature of noise in quantum systems interacting with an environment
at a finite temperature. With linear coupling to nonohmic baths or
at low temperatures, colored noise and nonlocal dissipation would appear
and for nonlinear coupling multiplicative noise is generally expected.
We derive a generalized fluctuation-
dissipation relation for these systems. Then using a model of quantum
Brownian motion in a bath of parametric oscillators, we show how noise
and dissipation can be related to the Bogolubov coefficients of
parametric amplification, which in the second-quantized sense,
depicts cosmological particle creation in a dynamic background.
We then calculate the influence functional and
study the noise characteristics of quantum fields as probed by
a particle detector. As examples, we show that an uniformly- accelerated
observer in flat space or an inertial observer in an exponentially
expanding (de Sitter) universe
would see a thermal particle spectrum, recovering the well-known results
of Unruh and Gibbons and Hawking
, as inspired by the Hawking effect in black holes.
We show how this method can be effectively used for treating the
backreaction of particle creation and other quantum field processes on the
dynamics of the early universe and black holes. We also discuss the
advantage of adopting the viewpoint of quantum open systems in addressing
some basic issues of semiclassical gravity and quantum cosmology.

\end{abstract}
\newpage

\setcounter{equation}{0}
\section{Introduction}

It is no secret that the message of this conference is `NOISE IS GOOD'.
In this talk I (BLH) want to
show that {\it not only is noise good, it is absolutely essential}. This is
because, as one school of cosmology believes and I think what every
cosmologist would have to confront ultimately when the questions of why and how
are insisted upon, noise, as fluctuations of the vacuum, is
possibly the only thing there was in the beginning  which, when ingeneously
construed can explain the birth of the universe and the germination of
everything in it, including spacetime and matter in their multifarious forms
and structures. \footnote {This general belief of the speaker (which may or may
not be shared by his collaborators) may echo
the more expressed views of the `birth of the universe from nothing'
scenario of Tryon, Brout, Englert, Gunzig, Spindel, Vilenkin, Linde and others
\cite{birth}, the `no-boundary' proposal of Hartle and Hawking \cite{HarHaw83},
the `free-lunch' advertisement of Guth's inflationary cosmology
\cite{inflation},
or the philosophical underpinnings of the
`austerity principle' and `it from bit'  quests of Wheeler \cite{Wheeler},
or the `gravity as metric elasticity' materialist worldview of Sakharov
\cite{Sakharov},
but the differences are probably greater than similarities when it comes to
comparing one's view on the universe.}

Noise from quantum fields play an active and sometimes decisive role in many
fundamental processes in cosmology and gravitation, especially near the
Planck time ($~10^{-43}$ sec from the Big Bang). This is the time when many of
us believe that the familiar concept of spacetime depicted by Einstein's
theory of general relativity emerged from an as-yet-unknown quantum theory of
gravity. Below this scale the world can be adequately described by
a semiclassical theory of gravity \cite{BirDav},
where quantized matter fields coexist with a classical spacetime. Many
qualitative changes are believed to have taken place at this energy scale,
amongst them the formation of spacetime depictable as a manifold, the
emergence of time, the creation of particle pairs from the vacuum,
the growth of fluctuations as seeds for galaxies, and  possible
phase transitions and the ensuing entropy generation.
It is also the cross-over point of
quantum to classical transition. It is for this reason that I think one can
indeed view {\it semiclassical gravity as a branch of mesoscopic physics},
the topic of the session to which this talk belongs.
The only difference is that instead of dealing with
the quantum to classical and micro- to macro- transition in the state of matter
we are dealing with the corresponding issues for spacetime and fields.

In two recent conference papers \cite{HuTsukuba,HuWaseda} I have described how
the concepts, viewpoints and techniques of non-equilibrium quantum
statistical mechanics can benefit the studies of
some basic issues in gravitation and cosmology, including quantum processes
in black holes and the early universe. Some examples are:

\noindent 1. Particle creation as parametric amplification of vacuum
fluctuations\\
2. Thermal radiance from accelerated observers and black holes
   as fluctuation-dissipation phenomena\\
3. Entropy generation from quantum stochastic and kinetic processes\\
4. Phase transitions in the early universe as  noise-induced processes\\
5. Galaxy formation from primordial quantum fluctuations\\
6. Anisotropy dissipation from particle creation as backreaction processes\\
7. Dissipation in quantum cosmology and the issue of the initial state\\
8. Decoherence, backreaction  and the semiclassical limit of quantum gravity\\
9. Stochastic spacetime and continuum limit, gravity as an effective theory\\
10. Topology change in spacetime and loss of quantum coherence problems\\
11. Gravitational entropy, singularity  and time asymmetry\\
12. `Birth' of the universe as a spacetime fluctuation and tunneling
phenomenon\\

Recently there is an increasing effort to understand these processes
in terms of statistical field theory (for a review, see \cite{HuBanff}).
Topics 5-8 are discussed in the references given at the end of this
introduction.
Some of these processes are not as well-understood as others. Indeed Topics
9-12 may not even be well-defined or posed. But there is hope that if one can
construct a more rigorous theory of noise in quantum fields in curved
spacetimes
one can begin to formulate these problems in a meaningful and solvable way.

Since this workshop is about noise and not about cosmology,
we will not discuss any cosmological issue here, but focus on the
theory of noise in quantum fields.  We will however for illustration purpose
draw in two basic processes in gravitation and cosmology where noise
manifests in a simple and direct way. The two processes are particle creation
and
detection (Topics 1,2 above). For background material on noise in open systems
see, e.g., \cite{LinWes,Gar,MosMcC}.

The starting point for the study of quantum fields is the vacuum,
that for statistical mechanics the fluctuations. If the ultimate task for
the cosmologist is to explain (better yet, to reconstruct) everything
from nothing, it is prudent to first  understand
the nature and behavior of noise in fields and spacetime.
Therefore, in this talk we will aim at addressing the following problems:\\

1) {\it How to define the characteristics of noise from a
coarse-grained environment.}
To prepare the stage for particle-field interaction and field-spacetime
coupling we study the quantum Brownian model (QBM) of a particle
interacting with a coupled oscillator bath.
(See e.g.,\cite{LinWes,Gar} for references to earlier work on this old topic).
To see the more general features of noise we consider the cases of nonlinear
coupling between the particle  and bath. We use the
Feynman-Vernon influence functional formalism
\cite{FeyVer,CalLeg83,Gra}  to show how to calculate the averaged effect
of the environment on the particle, here viewed as the system. In Sec. 2 we
show how all the statistical information of the bath (field) is
contained in the two kernels, the noise kernel and the dissipation
kernel, and we explain the physical meaning of these two kernels following
\cite{Zhang}.
The nonlocal kernels which appear at low temperature and for nonohmic baths
(which are already present for the bilinear coupling) lead to
colored noise. The nonlinear coupling leads to multiplicative noise.
One of us had earlier suggested that \cite{HuPhysica}
colored noise, rather than the familiar white noise, is expected to appear
commonly in cosmological and gravitational problems .
This was shown systematically in \cite{HPZ1,HPZ2,HuBelgium}.
Multiplicative noise in cosmology based on a nonlinear Langevin equation
of classical oscillator model is discussed in \cite{HabKan}.

2) {\it How to deduce a generalized fluctuation-dissipation relation \cite{fdr}
for quantum field systems}.  In the nonlinear QBM studied by Hu, Paz and Zhang
\cite{HPZ2}
the coupling of the particle to the bath is via a form
general in the system variable but of polynomial power in the bath variable.
This generalizes the bilinear coupling case studied by most researchers
previously.
They found a general fluctuation-dissipation relation (FDR) for this
class of models with nonlinear coupling [Eq. (3.8)].
The FD kernel has a temperature-dependent factor which varies with the
polynomial order ($k$) in the bath variables in the coupling. But at high
and zero temperatures this kernel becomes the same for linear and nonlinear
cases and the relation becomes insensitive to the coupling.
It is the belief of these authors that the FDR is a categorical
relation because it reflects on the self-consistency
in the interaction between the stochastic stimulation (noise and fluctuation)
from the environment and the averaged response of the system
(relaxation-dissipation) (cf. \cite{LinWes}). Later we will remark on
FDR for quantum fields in curved spacetime. (See point 5 below.)

3) {\it How to model the particle-field coupling with a parametric oscillator
bath} for the study of
non-equilibrium quantum field processes in a time-dependent background
spacetime.
We can extract the statistical information of a quantum field (like quantum
noise) by coupling a particle detector to it \cite{Unr}
and studying the detector's response to the fluctuations of the field.
We model the detector as a Brownian particle and the quantum field as
a bath of coupled oscillators with time dependent frequencies \cite{UnrZur}.
We assume in Sec. 4 and 5
a monopole detector coupled bilinearly to the field.
The normal modes of a quantum field obey an oscillator equation
with time-dependent natural frequencies. For flat space (Minkowski spacetime)
the time dependence of the modes is a simple sinusoidal, whereas for
dynamic spacetimes it has a more complex behavior which leads to
parametric amplification of vacuum fluctuations and backscattering of waves.
In the second-quantized formulation this  corresponds to particle creation
\cite{Par69,Zel70}. Cosmological particle creation is very strong at
the Planck time and its effect on the dynamics of the universe can be
very powerful \cite{cosbkr}.
The backreaction of these quantum field processes manifests as
dissipation effect, which is described by the dissipation kernel
\cite{CalHu87}.
We show how the influence functional can be expressed in terms of the Bogolubov
coefficients which appear in the unitary transformation between the
Fock space operators defined at different times. Since these coefficients
determine the amount of particles produced, {\it one can identify the origin of
noise in this system to particle creation} \cite{HM2,CalHuSG,HM3}.
On a related point, the transition of the system from quantum to classical
requires the diminishing of coherence in the wave function.
The noise kernel is found to be primarily responsible for this decoherence
process \cite{envdec,conhis}.
Decoherence can be studied by analyzing the magnitude of the diffusion
coefficients
in the master equation. We write down without derivation (see \cite{HM2}
for details) the master equation
for a QBM in a parametric oscillator bath (POB)
and indicate how it is different from
the case studied before with time-independent frequencies. This new result
is useful for the analysis of decoherence where parametric excitation
is present in the environment.
Here we aim not at the decoherence or the dissipation process,
but to focus on the very definition and nature of noise associated with
quantum fields in gravitation and cosmology.

4) {\it How do the characteristics of quantum noise vary
with the nature of the field}, the type of coupling between the field and the
background spacetime, and the time-dependence of the scale factor of the
universe.
As an example, in Sec. 5 we illustrate how a uniformly accelerating detector
observes a thermal spectrum. This way of understanding the Unruh
effect \cite{Unr} is recently discussed by Anglin \cite{Ang}.
As another example, we show that in a cosmology with an exponentially
expanding scale factor (the de Sitter universe), a thermal spectrum
is observed for a comoving observer (in the vacuum defined with respect
to the proper time). This effect was first proposed by
Gibbons and Hawking \cite{GibHaw} after the Hawking effect  for black holes
was discovered \cite{Haw75}. The viewpoint of quantum open systems
and the method of influcence functionals can, in our opinion,
lead to a deeper understanding of black hole thermodynamics and
quantum processes in the early universe \cite{HuWaseda}.
A fluctuation-dissipation
relation for quantum fields in black hole spacetimes was first suggested by
Sciama \cite{Sciama}, and later derived for de Sitter spacetime via linear
response
theory by Mottola \cite{Mottola}. These familiar cases all
deal with spacetimes with event horizons and thermal particle creation.
{}From earlier backreaction studies in semiclassical gravity
a general FDR was conjectured by one of us \cite{HuPhysica}
for quantum fields in  curved spacetimes
without event horizon. This corresponds to a non-equilibrium generalization
of the black hole case which we believe should capture the essense of the
particle creation backreaction processes in curved spacetime \cite{HuSin94}.

5) The backreaction of particle creation has been
studied in detail before \cite{cosbkr,CalHu87}. Indeed it was with
the aim of understanding the statistical meaning of dissipation
in this backreaction which led one of us to adopt
the influcence functional (or the equivalent coarse-grained Schwinger-Keldysh,
or closed-time-path effective action \cite{ctp})
method to problems in semiclassical gravity and cosmology
\cite{HuPhysica,HuTsukuba,HuWaseda}.
In Sec. 6 we outline a program for studying the backreaction of
particle creation in semiclassical cosmology. We use a model where the quantum
Brownian particle
and the oscillator bath are coupled parametrically. The field parameters
change in time through the time-dependence of the scale factor of the
universe, which is governed by the semiclassical Einstein equation.
We give an expression for the influence functional in terms of the
Bogolubov coefficients as a function of the scale factor. We indicate how one
can obtain a new, extended theory of semiclassical gravity which, in our
opinion,
is necessary for furthering the investigation of quantum and
statistical processes in curved spacetimes \cite{CalHuSG,HM3}.\\

This talk aims only as an introduction to the theory of noise in quantum
systems.
Its content is based mainly on two recent papers: Sec. 2, 3 on \cite{HPZ2}
and Sec. 4, 5 on \cite{HM2}.
The main theme of this talk, on the origin and nature of quantum
noise in gravitation and cosmology, is also discussed in  the following related
work from which the reader can find a wider exposition of topics:\\
1) On the galaxy formation problem
in inflationary cosmology \cite{HuBelgium,LafMat,Mat};\\
2) On noise and fluctuations in semiclassical gravity
\cite{CalHuSG,HM3,KuoFor};\\
3) On dissipation and initial conditions in quantum cosmology
\cite{disQC,SinHu,PazSin92}\\
4) On a fluctuation-dissipation theorem in cosmology,
  geometrodynamic noise and gravitational entropy \cite{HuWaseda,HuSin94}.

\newpage
\section{Quantum Noise from the Influence Functional}

Consider a Brownian particle interacting with a set of harmonic  oscillators.
The classical action of the Brownian particle is given by
\be
S[x]=\int_0^tds\Bigl\{{1\over 2}M\dot x^2-V(x)\Bigr\}.
\te
The action for the bath is given by
\be
S_b[\{q_n\}]
=\int_0^tds\sum_n\Bigl\{
 {1\over 2}m_n\dot q_n^2
-{1\over 2}m_n\omega^2_nq_n^2 \Bigr\}.
\te
We will assume in this and the next section that
the action for the system-environment interaction has the following form
\be
S_{int}[x,\{q_n\}]
=\int\limits_0^t ds\sum_n v_n(x) q_n^k  \label{int}
\te

\noindent where  $v_{n}(x) = -\lambda c_nf(x)$ and $\l$ is a
dimensionless coupling constant.
If one is interested only in the averaged effect of the environment on the
system
the appropriate object to study is
the reduced density matrix of the system $\r_r$, which is related to the
full density matrix $\r$ as follows
\be
\rho_r(x,x')
=\int\limits_{-\infty}^{+\infty}dq\int\limits_{-\infty}^{+\infty}
  dq'\rho(x,q;x',q')\delta(q-q').
\te
\noindent It is propagated in time by the propagator ${\cal J}_r$
\be
\rho_r(x,x',t)
=\int\limits_{-\infty}^{+\infty}dx_i\int\limits_{-\infty}^{+\infty}dx'_i~
 {\cal J}_r(x,x',t~|~x_i,x'_i,0)~\rho_r(x_i,x'_i,0~).
\te
If we assume that at a given
time $t=0$ the system and the environment are uncorrelated
\be
\hat\rho(0)=\hat\rho_s(0)\times\hat\rho_b(0),
\te
then ${\cal J}_r$ does not depend on the initial state of the system
and can be written as
\bea
{\cal J}_r(x_f,x'_f,t~|~x_i,x'_i,0)
& =& \int\limits_{x_i}^{x_f}Dx                     \label{prop}
   \int\limits_{x'_i}^{x'_f}Dx'~
   \exp{i\over \hbar}\Bigl\{S[x]-S[x']\Bigr\}~{\cal F}[x,x']  \nn \\
& =& \int\limits_{x_i}^{x_f}Dx
   \int\limits_{x'_i}^{x'_f}Dx'~
   \exp{i\over \hbar} {\cal A}[x,x']
\tea
where the subscripts  $i,f$ denote initial and  final variables,
and ${\cal A}[x,x']$ is the effective action for the open quantum system.
The influence functional ${\cal F}[x,x']$ is defined as
\bea
 {\cal F}[x,x'] & = &\int\limits_{-\infty}^{+\infty}dq_f
 \int\limits_{-\infty}^{+\infty}dq_i
 \int\limits_{-\infty}^{+\infty}dq'_i          \label{ifbm}
 \int\limits_{q_i}^{q_f}Dq
  \int\limits_{q'_i}^{q_f}Dq' \nn \\
& & \times \exp{i\over\hbar}\Bigl\{
  S_b[q]+S_{int}[x,q]-S_b[q']-S_{int}[x',q'] \Bigr\}
  \rho_b(q_i,q'_i,0) \nn \\
& = & \exp{i\over\hbar} \delta {\cal A}[x,x']
\tea
where $\delta {\cal A}[x,x']$ is the influence action.  Thus $ {\cal A}[x,x'] =
S[x]-S[x'] + \delta {\cal A}[x,x']$.

{}From its definition it is obvious that if the interaction term is zero, the
influence functional is equal to unity and the influence action is zero.
In general, the influence functional
is a highly non--local object. Not only does it depend on the time history,
but --and this is the more important property-- it also
irreducibly mixes the two sets
of histories in the path integral of (2.7).
Note that the histories
$ x $ and $ x' $ could be interpreted as moving
forward and backward in time respectively.
Viewed in this way, one can see the similarity of the influence functional
and the generating functional in the closed-time-path, or Schwinger-Keldysh
\cite{ctp} integral formalism.

It can be shown \cite{HPZ2} that the influence action for the model given
by the interaction in (2.3)  to second order in $\lambda$ is given
by
\bea
& &\delta {\cal A} [x,x']
= \Bigl\{\int\limits_0^tds~[-\delta V(x)~]
  -\int\limits_0^tds~[-\delta V(x')]~\Bigr\} \nn \\
& & -\int\limits_0^tds_1\int\limits_0^{s_1}ds_2~\lambda^2 \label{iabm}
   \Bigl[f(x(s_1))-f(x'(s_1))\Bigr]\mu^{(k)}(s_1-s_2)
   \Bigl[f(x(s_2))+f(x'(s_2))\Bigr]  \nn \\
& &+i\int\limits_0^tds_1\int\limits_0^{s_1}ds_2~\lambda^2
   \Bigl[f(x(s_1))-f(x'(s_1))\Bigr]\nu^{(k)} (s_1-s_2)
   \Bigl[f(x(s_2))-f(x'(s_2))\Bigr]
\tea
where $\delta V(x)$ is a renormalization of the potential that arises
from the contribution of the bath. It appears only for even $k$ couplings.
For the case $k=1$ the above result is exact.
This is a generalization of the
result obtained in \cite{FeyVer} where it was shown that the
non-local kernel $\mu^{(k)}(s_1-s_2)$ is associated with dissipation or
the generalized viscosity function that appears in the corresponding
Langevin equation and $\nu^{(k)}(s_1-s_2)$ is associated with the
time correlation function of the stochastic noise term. The dissipation
part has been studied in detail  by
Calzetta, Hu, Paz, Sinha and others \cite{CalHu87,disQC,SinHu,PazSin92,CalHuSG}
in cosmological backreaction problems. We shall only
discuss the noise part of the problem. In general $\nu$ is nonlocal, which
gives rise to colored noises.
Only at high temperatures would the noise kernel become a delta function,
which corresponds to a white noise source. Let us see the meaning
of the noise kernel.

The noise part of the influence functional is given by
\be
\exp\{ -{1\over \hbar}\int\limits_0^tds_1\int\limits_0^{s_1}ds_2~
   \Bigl[f(x(s_1))-f(x'(s_1))\Bigr]\nu^{(k)}(s_1-s_2) \label{expnoise}
   \Bigl[f(x(s_2))-f(x'(s_2))\Bigr]\}
\te
where $\nu$ is redefined by absorbing the $\l^2$.
This term can be rewritten using the following functional Gaussian identity
\cite{FeyVer}
which states that the above expression (2.10) is equal to
\be
\int {\cal D}\xi^{(k)}(t) {\cal P}[\xi^{(k)}]\exp\{ {i\over
\hbar}\int\limits_0^t
ds \xi^{(k)}(s)[f(x(s)) - f(x'(s))]\}
\te
where
\be
{\cal P}[\xi^{(k)}] = N^{(k)} \exp\{-{1\over        \label{noisedist}
\hbar}\int\limits_0^tds_1\int\limits_0^tds_2 \ha \xi^{(k)}(s_1)
[\nu^{(k)} (s_1 - s_2)]^{-1}\xi^{(k)}(s_2) \}
\te
is the functional distribution of $\xi^{(k)}(s)$ and $N^{(k)}$ is a
normalization factor given by
\be
[N^{(k)}]^{-1} = \int{\cal D}\xi^{(k)}(s)\exp\{-{1\over \hbar}
\int\limits_0^tds_1\int\limits_0^tds_2\xi^{(k)}(s_1)[\nu^{(k)}(s_1-s_2)]^{-1}
\xi^{(k)}(s_2)\}.
\te
The influence functional can then be rewritten as
\bea
{\cal F}[x,x'] &=& \int{\cal D}\xi^{(k)}(s){\cal P}[\xi^{(k)}] exp{i\over
\hbar}\delta
{\cal A}[x,x',\xi^{(k)}]\nn \\
&\equiv& {\left\langle \exp {i\over \hbar}\delta {\cal
A}[x,x',\xi^{(k)}]\right\rangle}_{\xi}
\tea
where
\bea
& &\delta {\cal A}[x,x',\xi^{(k)}]  =   \int\limits_0^tds~\Bigl\{-\delta
V(x)~\Bigr\}
  -\int\limits_0^tds~\Bigl\{-\delta V(x')~\Bigr\} \nn \\
& & -\int\limits_0^tds_1\int\limits_0^{s_1}ds_2~
   \Bigl[f(x(s_1))-f(x'(s_1))\Bigr]\mu^{(k)}(s_1-s_2)  \label{ianoise}
   \Bigl[f(x(s_2))+f(x'(s_2))\Bigr]  \nn \\
& & - \int\limits_0^t ds \xi^{(k)}(s)f(x(s))
+\int\limits_0^tds\xi^{(k)}(s)f(x'(s))
\tea
so that the reduced density matrix can be rewritten as
\be
\rho_r(x,x') = \int{\cal D}\xi^{(k)}(s){\cal
P}[\xi^{(k)}]\rho_r(x,x',[\xi^{(k)}]).
\te
{}From equation (2.15) we can view $\xi^{(k)}(s)$ as a
nonlinear external stochastic force and the
reduced density matrix is calculated by taking a stochastic average
over the distribution $P[\xi^{(k)}]$ of this source.

This is a Gaussian type noise since from (2.12), we can see
that the distribution functional is Gaussian. The noise is therefore
completely characterized by
\bea
{\langle\xi^{(k)}(s)\rangle}_{\xi^{(k)}} &=& 0 \nn\\
{\langle\xi^{(k)}(s_1)\xi^{(k)}(s_2)\rangle} &=& \hbar\nu^{(k)}(s_1-s_2).
\tea
We see that the non-local kernel $\nu^{(k)}(s_1-s_2)$ is just the two
point time correlation function of the external stochastic source
$\xi^{(k)}(s)$ multiplied by $\hbar$.

In this framework, the expectation value of any quantum mechanical
variable $Q(x)$ is given by \cite{Zhang}
\bea
\langle Q(x)\rangle & = & \int{\cal D}\xi^{(k)}(s){\cal
P}[\xi^{(k)}]\int\limits_{-\infty}
^{+\infty}dx \rho_r(x,x,[\xi^{(k)}])Q(x) \nn \\
& = & {\left\langle {\langle Q(x)\rangle}_{quantum}\right\rangle}_{noise}.
\tea
This summarizes the interpretation of $\nu^{(k)}(s_1-s_2)$ as a noise or
fluctuation kernel.

We will now derive the semiclassical equation of motion generated by the
influence action (2.9). This will allow us to
see why the
kernel $\mu^{(k)}(s_1-s_2)$ should be associated with dissipation.
Define a ``center-of-mass" coordinate $\Sigma$ and a
``relative" coordinate $\Delta$ as follows
\bea
\Sigma(s) & = & {1\over 2}[x(s) + x'(s)] \nn\\
\Delta(s) & = & x'(s) - x(s).
\tea
The semiclassical equation of motion for $\Sigma$ is derived by demanding
(cf. \cite{CalHu87})
\be
\frac{\delta}{\delta\Delta}\Bigl[S[x]-S[x']+\delta {\cal
A}[x,x']\Bigr]\bigg|_{\Delta=0}=0.
\te
Using the sum and difference coordinates (2.19) and the influence action
(2.9) we find that (2.20) leads to
\be
\frac{\partial L_r}{\partial \Sigma}-\frac{d}{dt}\frac{\partial L_r}{\partial
\dot{\Sigma}} - 2\frac{\partial f(\Sigma)}{\partial \Sigma}\int\limits_0^t
ds~\gamma^{(k)}(t-s)\frac{\partial f(\Sigma)}{\partial \Sigma}\dot{\Sigma}=
F_{\xi^{(k)}}(t)
\te
where $\frac{d}{ds}\gamma^{(k)}(t-s)=\mu^{(k)}(t-s)$.
We see that this is in the form of
a classical Langevin equation with a nonlinear stochastic force
$F_{\xi^{(k)}}(s) = -\xi^{(k)}(s) \frac{\partial f(\Sigma)}{\partial \Sigma}$.
This corresponds to multiplicitive noise unless $f(\Sigma)=\Sigma$ in which
case it is
additive. $L_r$ denotes a renormalised
system Lagrangian. This is obtained by absorbing a surface term and the
potential renormalisation
in the influence action into the system action. The
nonlocal kernel $\gamma^{(k)}(t-s)$ is responsible for non-local dissipation.
In special cases like a high temperature ohmic environment,
this kernel becomes a delta function and hence the dissipation is local.

\section{Fluctuation-Dissipation Relation for Systems with Colored and
Multiplicative Noise}

Recall that the label $k$ is the order of the bath variable to which the system
variable is coupled
to. $\gamma^{(k)}(s)$ can be written as a sum of various contributions
\be
\gamma^{(k)}(s)=\sum_l\gamma^{(k)}_l(s)
\te
where the sum is over even (odd) values of $l$ when $k$ is even (odd).
To derive the explicit forms of each dissipation kernel, it is useful to
define first the spectral density functions
\be
I^{(k)}(\omega)
= \sum_n~\delta(\omega-\omega_n)~k~\pi\hbar^{k-2}~
   {\lambda^2 c_n^2(\omega_n)\over (2m_n\omega_n)^k }.
\te
It contains the information about the environmental mode density and coupling
strength as a function of frequency.
Different environments are classified according to the
functional form of the spectral density $I(\o)$.  [On physical grounds, one
expects the spectral density to go to zero for very high frequencies, and thus
a certain cutoff frequency $\Lambda$  (which is a property of the
environment) is often introduced
such that $I(\o)\rightarrow 0$ for $\omega>\Lambda$.]
The environment is
classified as ohmic \cite{CalLeg85,Gra,UnrZur,HPZ1}
if in the physical range of frequencies
($\omega<\Lambda$) the spectral density is such that
$I(\o)\sim\omega$, as supra-ohmic if $I(\o)\sim\omega^n , n>1$
or as sub-ohmic if $n<1$. The most studied ohmic case corresponds to an
environment which induces a dissipative force linear in the velocity of the
system.

In terms of these functions, the dissipation kernels can be written
as
\be
\gamma^{(k)}_l(s)
 =\int\limits_0^{+\infty}{d\omega\over\pi}
 ~{1\over \omega}I^{(k)} (\omega)
 ~ M^{(k)}_l(z)~\cos l\omega s
\te
where $M^{(k)}_l(z)$ are temperature dependent factors derived in \cite{HPZ2}.
Analogously, the noise kernels $\nu^{(k)}(s)$
can also be written as a sum of various contributions
\be
\nu^{(k)}(s) = \sum_l \nu^{(k)}_l(s)
\te
\noindent where the sum runs again over even (odd) values of $l$ for
$k$ even (odd). The kernels $\nu^{(k)}_l(s)$ can be written as
\be
\nu^{(k)}_l
=\hbar\int\limits_0^{+\infty}{d\omega\over\pi}~
 I^{(k)} (\omega)~N^{(k)}_l(z)~\cos l\omega s
\te
where $ N^{(k)}_l (z)$ is another set of temperature- dependent factors given
by
\cite{HPZ2}

To understand the physical meaning of the noise kernels of different orders,
we can think of them as being associated with  $l$ independent stochastic
sources that are coupled to the Brownian particle through interaction
terms of the form [Eq.(2.15)]
\be
\int\limits_0^tds~\sum_l~\xi_l^{(k)}(s)~f(x).
\te
\noindent This type of coupling generates a stochastic force in the associated
Langevin equation
\be
F_{\xi_l^{(k)}}(s)=-\xi_l^{(k)}(s)\frac{\partial f(x)}{\partial x}
\te
\noindent which corresponds to multiplicative noise.
The stochastic sources $\xi_l^{(k)}$ have a probability distribution given by
(2.12)
which generates the correlation functions (2.17) for each $k$ and $l$.

To every stochastic source we can associate a dissipative term that is
present in the real part of the influence action. The dissipative and the
noise
kernels are related by generalized fluctuation--dissipation relations
of the following form
\be
\nu^{(k)}_l(t)
=\int\limits_{-\infty}^{+\infty}ds~K^{(k)}_l(t-s)~\gamma^{(k)}_l(s)
\te
\noindent where the kernel $ K^{(k)}_l(s) $ is
\be
K^{(k)}_l(s)
=\int\limits_0^{+\infty}{d\omega\over\pi}~
  L^{(k)}_l(z)~l~\omega~\cos~l\omega s
\te
and the temperature-dependent factor $L^{(k)}_l(z)= N^{(k)}_l(z)/
M^{(k)}_l(z)$.

A fluctuation dissipation relation of the form (3.8) exists for
the linear case where the temperature dependent factor appearing in (3.9)
is simply $L^{(1)}=z$. The fluctuation-dissipation kernels $K_l^{(k)}$
have rather complicated forms except in some special cases.
In the high temperature limit, which is characterized
by the condition
$ k_BT\gg \hbar\Lambda $, where $ \Lambda $ is the cutoff frequency of
the environment, $z=\coth \beta\hbar\omega/2
\to 2/\beta\hbar\omega$                                      
we obtain
\be
L^{(k)}_l(z) \to {{2k_BT}\over {\hbar\omega}}.
\te
\noindent In the limit $ \Lambda \to +\infty $, we get the general result
\be
K^{(k)}_l(s)= {2k_BT\over \hbar}\delta(s)
\te
\noindent which tells us that at high temperature there is
only one form of fluctuation-dissipation relation, the Green-Kubo relation
\cite{fdr}
\be
 \nu^{(k)}_l(s)
={2k_BT\over \hbar}\gamma^{(k)}_l(s).
\te
\noindent In the zero temperature limit, characterized by $~ z \to 1,~ $
we have
\be
L^{(k)}_l(z) \to l.
\te
\noindent The fluctuation-dissipation kernel becomes $k$-independent
and hence identical to the one for the linearly- coupled case
\be
K(s)
=\int\limits_0^{+\infty}
 {d\omega\over\pi}~\omega\cos\omega s.
\te

It is
interesting to note that the fluctuation-dissipation relations for the
linear and the nonlinear dissipation models are exactly identical both
in the high temperature and in the zero temperature limits. In other words,
they are not very sensitive to the different
system-bath couplings at both high and zero temperature limits.
The fluctuation-dissipation relation reflects a
categorical relation (backreaction) between the stochastic stimulation
(fluctuation-noise) of the environment and the averaged response of a system
(dissipation) which has a much deeper and universal meaning than that
manifested
in specific cases or under special conditions.
\footnote{A given environment is characterized by the spectral densities
$I^{(k)}(\omega)$ and it is clear that if these functions are appropriately
chosen, the form of the noise and dissipation kernels can be simplified
considerably. For example, if the spectral density is
$I^{(k)}(\omega)\sim\omega^k$,
the noise and the dissipation kernels become local kernels
in the high temperature limit. In that case we have
$\gamma^{(k)}_l(s)\sim (k_BT)^{k-1}\delta(s),~~
\nu^{(k)}_l (s)\sim(k_BT)^k\delta(s)$.
Note that $\gamma_l^{(k)}$ depends upon the temperature and will produce a
temperature dependent friction term in the effective equations of motion.
On the other hand if the spectral density is the same linear function for
all $(k)$, i.e., $I^{(k)}(\omega)\sim\omega$,
the dissipation kernel will become local in the low temperature limit:
$\gamma^{(k)} \sim \delta (s)$,
but the noise remains colored
due to the nontrivial fluctuation dissipation relation (3.8).
However, as we will show below, for quantum fields the spectral density
are fixed by their own character and cannot be adjusted arbitrarily.}

\section{Brownian Particle in a Bath of Parametric Oscillators}

The previous two sections showed how noise and dissipation are generated
using general system environment couplings within the
quantum Brownian motion paradigm. This extension
of the quantum Brownian motion paradigm to non-linear couplings and
nonlocal noise and dissipation is essential if we are
to address the issues in cosmology and gravity outlined in the Introduction.
Since the early universe is rapidly expanding we need a formalism which allows
us to study the non-equilibrium
quantum statistical processes in time- dependent backgrounds.
In the following two sections we will discuss the Brownian motion of a
quantum particle in a bath of parametric oscillators and show how this model
can be used to treat particle creation and detection processes in the
early universe and black holes. To lessen the complexity of the problem
we will consider only linear coupling between the system and the bath.

Consider now the system being a parametric oscillator with mass $M(s)$,
cross term $B(s)$ and natural (bare) frequency $\Omega(s)$.
The environment is also modelled by a set of parametric oscillators with mass
$m_n(s)$, cross term $b_n(s)$ and natural frequency $\omega_n(s)$.
We assume that the system-environment coupling is linear in the
environment coordinates with strength $c_n(s)$, but general in the system
coordinate. The action of the combined
system $+$ environment is
\bea
S[x,q] &=&S[x]+S_E[q]+S_{int}[x,q] \nn \\
& =&\int\limits_0^tds \Biggl[ {1\over 2}
  M(s)\Bigl( \dot x^2+B(s)x\dot{x}-\Omega^2(s)x^2\Bigr) \nn \\
& + &\sum_n\Bigl\{  {1\over 2}m_n(s)
  \Bigl(\dot q_n^2+b_n(s)q_n\dot{q}_n- \omega^2_n(s)q_n^2\Bigr) \Bigr\}
  + \sum_n\Bigl(-c_n(s)f(x)q_n\Bigr)\Biggr]
\tea
\noindent where $x$ and $q_n$ are the coordinates of the particle and the
oscillators respectively.

\subsection{Bogolubov Transformation and Particle Creation}

All the information about the quantum dynamics of the bath
parametric oscillators are contained in the two complex numbers,
$\a$ and $\b$, known as the Bogolubov coefficients.
They obey two coupled first order equations \cite{Par69,Zel70}
\begin{eqnarray}
\dot{\alpha} & = & -ig^*\beta-ih\alpha \nn \\
\dot{\beta} & = & ih\beta+ig\alpha
\end{eqnarray}
and are related by the Wronskian condition $|\a|^2 - |\b|^2 = 1$.
The time-dependent coefficients are given by
\begin{equation}
g(t)=\frac{1}{2}\left(\frac{m(t)\omega^2(t)}{\kappa}
+\frac{m(t)b^2(t)}{4\kappa}-\frac{\kappa}{m(t)}+ib(t)\right)
\end{equation}
\begin{equation}
h(t)=\frac{1}{2}\left(\frac{\kappa}{m(t)}+\frac{m(t)\omega^2(t)}{\kappa}
+\frac{m(t)b^2(t)}{4\kappa}\right).
\end{equation}
where $\kappa$ is an arbitrary positive real constant that is usually chosen so
that
$g=0$ at $t_i$.
Thus if $b_n=0$ we will usually have $\kappa=m(t_i)\o(t_i)$.
Given the initial condition for the propagator,
Eq.(4.2) must satisfy the initial conditions $\alpha(t_i)=1, \beta(t_i)=0$.
In a cosmological background, the time dependence of
$g$ and $h$ are parametric in nature, i.e., it comes from the time-dependent
scale factor $a$.

One can use the squeeze state language to depict particle creation
\cite{GriSid,HKM}. The unitary evolution operator $\hat{U}$ for this
time-dependent
system
can be expressed as a product of the squeeze and rotation operators
$\hat S, \hat R$ \cite{sqst}
\begin{equation}
\hat{U}(t,t_i)=\hat{S}(r,\phi)\hat{R}(\theta)
\end{equation}
where
\begin{equation}
\hat{R}(\theta)=e^{-i\theta \hat{B}},\;\;\;\;\;
\hat{S}(r,\phi)=\exp [r(\hat{A}e^{-2i\phi}-\hat{A}^{\dag}e^{2i\phi})].
\end{equation}
Here
\be
\hat{A} = \hat{a}^2/2, ~~~ \hat{B}=\hat{a}^{\dag}\hat{a}+1/2
\te
and $\hat{a}, \hat{a}^\dagger$ are the annihilation and creation operators of
the second-quantized modes.
The Bogolubov coefficients $\alpha$ and $\beta$ are related to the three
real parameters, $r$, the squeeze parameter, $\phi$, the squeeze angle,
and $\theta$, the rotation angle by
\begin{equation}
\alpha=e^{-i\theta}\cosh r,\;\;\;\;\;\beta=-e^{-i(2\phi+\theta)}\sinh r.
\end{equation}

The exact influence action for the model (4.1) takes the form (2.9) with
$\delta V(x)=0$ and $k=\l=1$.
For an initial thermal state \footnote{In Ref. \cite{HM2} we consider a
squeezed
thermal initial state which is a generalisation of the initial state used here.
This form of initial state is of interest in quantum optics.},
the dissipation and noise kernels are calculated to be \cite{HM2}
\be
\mu(s,s')=\frac{i}{2}\int_0^{\infty}d\omega I(\omega,s,s')
[X^* X' - X X'^*]
\te
\be
\nu(s,s')  =  \frac{1}{2}\int_0^{\infty}d\omega I(\omega,s,s')
\coth\left(\frac{\hbar\omega(t_i)}{2k_BT}\right)[X^* X' + X X'^*]
\te
where $X \equiv X_\o (s) \equiv \a_\o(s) + \b_\o(s)$ and $X' \equiv X_\o(s')$.
\footnote{If we assume $b=0$ and $m=1$ we can show using  (4.2) that
$X$ is a solution of
$\ddot{X}+\omega^2_n(t)X=0$
subject to the boundary condition $X(t_i)=1$. From this one can show that
with a thermal initial state in the high temperature limit our quantum theory
gives the correct classical result \cite{HabKan}.}
The spectral density, $I(\o,s,s')$ defined by
\be
I(\omega,s,s')=\sum_n\delta(\omega-\omega_n)\frac{c_n(s)c_n(s')}{2\kappa}
\te
is obtained in the continuum limit. It
contains information about the environmental mode density and coupling
strength as a function of frequency.
We see from (4.9-10) that the effect of parametric amplification in the bath
affects both the noise and dissipation kernels.

{}From (2.20) with the influence action (2.9) (with $\delta V(x)=0$ and
$k=\l=1$),
we find that the semiclassical equation is given by
\be
\frac{\partial L_r}{\partial x}-\frac{d}{dt}\frac{\partial L_r}{\partial
\dot{x}}
- 2\frac{\partial f(x)}{\partial x}
\int\limits_0^t ds~\gamma(t,s) \frac {\partial f(x)}{\partial x}\dot{x}
= -\frac{\partial f(x)}{\partial x} \xi(t)
\te
where $\xi$ is a zero mean gaussian stochastic force with the correlator
$\langle \xi(t)\xi(t')\rangle=\hbar\nu(t,t')$ and $L_r$ is an arbitrary system
Lagrangian renormalised by a surface term from the nonlocal kernel. The noise
and dissipation kernels are no longer stationary due to the time dependence of
the bath.

\subsection{Noise and Decoherence}

Finding how the classical features arise from a quantum system is a fundamental
issue for all physical systems, including the description of the universe
itself \cite{HarMisfest}.
In many cosmological processes it is essential to be able to say when
some degree of freedom has become effectively classical. Only when this happens
will the semiclassical description will appropriate.\footnote {For example,
in the
inflationary universe it seems plausible that the amplification of quantum
fluctuations of a scalar field can act as seeds for primordial density
perturbations. In this case the transition from quantum to classical
fluctuations is critical since it is responsible for breaking
the spatial translational invariance of the vacuum,
which is necessary in order to generate primordial density perturbations
\cite{BraLaf,LafMat}. The quantum  origin and nature of noise in the
generation of structure in the primordial universe is discussed in
\cite{HuBelgium}.}
To understand the quantum to classical transition an essential factor
is the suppression of the interference terms in the system of interest.
This can be achieved by letting the system interact with a coarse-grained
environment \cite{envdec}. (The problem is better formulated in terms of
decoherent or consistent histories \cite{conhis}.) Noise in the environment
here plays two important roles: one in decohering the system and causing it
to assume a classical character, the other in imparting a dissipative
behavior in the system dynamics. These two processes, decoherence and
dissipation, usually occur at very different time scales. Decoherence
is effectively achieved when thermal fluctuations overtake the vacuum
fluctuations \cite{HuZhaUncer,AndHal}. For macroscopic objects
at high temperatures, decoherence time is much faster than relaxation time.
This can be studied by analyzing the relative importance of the respective
terms in the master equation for the quantum open system.

A full quantum mechanical description of the dynamics of the open system
is given by the  propagator ${\cal J}_r$ of the reduced density matrix (2.7),
which can be, and has been derived exactly for the bilinear coupling case
\cite{FeyVer,CalLeg83,Gra,HPZ1,HM2}.
Using this we can derive the master equation
for the reduced density matrix.
The master equation is useful
because it separates out the different non-equilibrium  quantum processes
generated by the bath on the system.

The exact master equation for a system interacting with a bath described
by a general time-dependent quadratic Hamiltonian in a squeezed thermal initial
state is derived to be \cite{HM2}
$$
\eqalign{
&i\hbar{\partial\over\partial t}~\rho_r(x,x',t)
=\biggl\{ \biggl[  -{\hbar^2\over {2M(t)}}\Bigl({\partial^2\over\partial x^2}-
  {\partial^2\over\partial x'^2}\Bigr)
 +\frac{M(t)}{2}\Omega^2_{ren}(t,t_i)\bigl(x^2 -x'^2 \bigr)
 \biggr]~\cr
& -i\hbar\Gamma(t,t_i)(x-x')\Bigl({\partial\over\partial x}
  -{\partial\over\partial x'}\Bigr)~
 +iD_{pp}(t,t_i)(x-x')^2~ \cr
& -
\hbar\Bigl(D_{xp}(t,t_i)+D_{px}(t,t_i)\Bigr)(x-x')\Bigl({\partial\over\partial
x}
  +{\partial\over\partial x'}\Bigr)
 -i\hbar^2 D_{xx}(t,t_i)\frac{\partial^2}{(\partial x+\partial x')^2} \biggr\}
 \rho_r(x,x',t) \cr}
$$
where $\Omega_{ren}$ is the renormalized frequency, $\Gamma$ is the dissipation
coefficient and the $D's$ are the diffusion coefficients.
\footnote{
The master equation differs from that of the time-independent oscillator bath
\cite{HPZ1} by more than changing the kernels.
There the dissipation kernel is stationary (i.e a function of $(s-s')$) and
$D_{xx}$ is absent. A non-stationary factor
enters in all the diffusion coefficients and $D_{xx}$ depends solely on it.}
In operator form, it reads
\bea
i\hbar\frac{\partial}{\partial t}\hat{\rho}_r (t)&= & [\hat{H}_{ren},
\hat{\rho}]+iD_{pp}(t)[\hat{x},[\hat{x},\hat{\rho}]]+iD_{xx}(t)[\hat{p},[\hat{p},
\hat{\rho}]] \nn \\
& +&
iD_{xp}(t)[\hat{x},[\hat{p},\hat{\rho}]]+iD_{px}(t)[\hat{p},[\hat{x},\hat{\rho}]]
+\Gamma(t) [\hat{x},\{\hat{p},\hat{\rho}\}]
\tea
where
\be
\hat{H}_{ren}=\frac{\hat{p}^2}{2M}+\frac{M}{2}\O_{ren}(t)\hat{x}^2.
\te

The first term is the Liouville or streaming term describing the free unitary
dynamics of the system but with a renormalised
frequency. The last term proportional to $\Gamma$ is responsible for
dissipation. The renormalised frequency and the dissipation
coefficient $\Gamma$ do not depend on the noise kernal (as they
depend on the environment only via the functions $u_i$ defined in
\cite{HPZ1,HPZ2,HM2}.
The terms proportional to $D_{xx},D_{pp}, D_{xp}$ and $D_{px}$ generate
diffusion in the
variables $p^2, x^2$ and $ xp+px$ respectively. This can be seen by going
from the master equation to the Fokker-Planck equation
for the Wigner function \cite{Dekker}. The diffusion coefficients
are affected by both noise and dissipation kernels (as they
depend both on $u_i$ and $a_{ij}$). From the
master equation we know that $D_{xx}$ and $D_{pp}$ generate decoherence in $p$
and $x$ respectively. Decoherence is a critical process for the quantum
to classical transition. It is also responsible for entropy generation
and other quantum statistical effects.

\section{Particle-Field Interaction}

Let us now examine how to define the noise of a quantum field, both in
flat and curved spacetimes. It is easy to show that a field can be represented
as a parametric bath of oscillators \cite{BirDav}.
To study the noise properties of a quantum field, we  introduce
a particle detector and assume some interaction, the simplest being a monopole
linear coupling.
We consider two cases: an accelerated detector in flat space, and
an inertial detector in an expanding universe. It is seen that
if the acceleration is uniform or if the scale factor undergoes an exponential
expansion then the noise observed in the detector's proper time is thermal.
These two examples were first given in 1976 and 1977 by Unruh \cite{Unr} and
Gibbons and Hawking \cite{GibHaw} respectively. We use these well-known
examples to illustrate the
physics of the problem and to demonstrate the power of the
influence functional formalism in extracting the statistical information
of the system and bath.

\subsection{Accelerated Observer}

We consider a massive scalar field $\Phi$ in a two- dimensional
flat space with mode decomposition
\begin{equation}
\Phi(x)=\sqrt{\frac{2}{L}}\sum_{k}[q_{k}^+
\cos kx + q_{k}^- \sin kx].
\end{equation}
The Lagrangian for the field can be expressed as a sum of oscillators
with amplitudes $q^{\pm}_k$ for each mode
\be
L (s)=\frac{1}{2}\sum_{\sigma}^{+-}\sum_{k}
\left[(\dot{q}_{k}^{\sigma})^2
-\o_k^2 q_k^{\sigma2}\right].
\end{equation}
This corresponds to the case in  (4.1) with $m_n=1, b_n=0$.
Since $\o_k^2 = (k^2+m^2)$ in flat space is time-independent,
$ \a=e^{-i\o t},\;\;\b=0 $,
where $\a=1$ initially $t=0$. Substituting these into (4.9-10)
one obtains for an inertial detector in a thermal bath
\be
\mu(s,s')=-\int_0^{\infty}dk~ I(k)\sin \o(s-s')
\te
and
\be
\nu(s,s')=\int_0^{\infty}dk~\coth\left(\frac{\hbar \o}{2k_BT}\right)I(k)
\cos\o(s-s').
\te

Now consider an observer undergoing constant acceleration $a$ in this field
with trajectory
\be
x(\t)=\frac{1}{a}\cosh a\t,\;\;\;s(\t)=\frac{1}{a}\sinh a\t.
\te
We want to show that the observer detects thermal radiation.
Let us first find the spectral density. For a monopole detector the
particle- field interaction is
\be
{\cal L}_{int}(x)=-\e r\Phi(x)\delta(x(\t)).
\te
where they are coupled at the spatial point $x(\t)$ with coupling strength
$\e$
and $r$ is the detector's internal coordinate.
Integrating out the spatial variables we find that
\be
L_{int}(\t)=\int{\cal L}_{int}(x)dx=-\e r\Phi (x(\t)).
\te
Using (5.1-2) we see that the accelerated observer is coupled
to the field with  effective coupling constants
\be
c_n^+(s)=\e \sqrt{\frac{2}{L}}\cos kx(\t),\;\;c_n^-(s)=\e
\sqrt{\frac{2}{L}}\sin kx(\t).
\te
With this we find from (4.11) that in the continuum limit (using
$\sum_k\rightarrow\frac{L}{2\pi}\int dk,\;\kappa =\omega$)
\be
I(k)=\frac{\e^2}{2\pi\o}\cos k[x(\t)-x(\t')].
\te

The expressions for the noise and dissipation kernels in a
 zero-temperature field (assummed in an initial vacuum state)
can be obtained from
the above result for finite temperature bath by setting
$z \equiv \coth \b \hbar \o /2 \rightarrow 1$. We have
\be
\zeta(s(\t),s(\t'))=\nu(s,s')+i\mu(s,s')=\int_0^{\infty}dk~I(k)e^{-i\omega[s(\t)-s(\t')]}.
\te
Now putting in the spectral density function $I(k)$ we get
\be
\zeta(s(\t),s(\t'))=\frac{\e ^2}{4\pi}\int_{-\infty}^{\infty}\frac{dk}{\omega}
e^{-ik[x(\t)-x(\t')]-i\omega[s(\t)-s(\t')]}.
\te
We can write this as  \cite{Ang}
\be
\zeta(\t,\t')=\int
\frac{d\omega}{\omega}G(a,\omega)\Bigl[\coth(\pi\omega/a)\cos
\omega(\t-\t')-i\sin \omega(\t-\t')\Bigr]
\te
where
\be
G(a,\omega)=\frac{\e ^2\omega}{a\pi^2}\sinh (\pi\omega/a)[K_{i\omega/a}(m/a)]^2
\te
and $K$ is the Bessel function. Comparing this with (5.4) we see that
a thermal spectrum is detected by
a uniformly-accelerating observer at temperature
\be
k_B T= {a \over {2 \pi}}.
\te
This was first found by Unruh and presented in this form recently
by Anglin \cite{Ang}. Note that there is a trade-off between the thermal
effect of the bath as detected by an inertial observer and the kinematic
effect of the detector undergoing uniform acceleration in a vacuum.
Although it is well-known that a uniformly accelerating observer sees
an exact thermal radiance, the case of arbitrary motion is perhaps lesser
known. It is certainly more difficult to analyze if one interprets the
Unruh effect (or Hawking effect for black holes) in the geometric sense,
i. e., via the event horizon, which does not exist for all times in
this more general case.
In the statistical mechanical viewpoint we are espousing, it is easier
to understand that noise is always present no matter how the detector moves.
Indeed this formalism
tells one how to calculate the form of noise for an arbitrary particle
trajectory
and field.
It also separates the kinematic and the thermal effects so one can interpret
the physics clearly.

\subsection{Thermal Radiance in de Sitter Space}

Consider now the 4-dimensional Robertson-Walker (RW) spacetime with line
element
\be
ds^2= dt^2 -\sum_i a^2(t)dx_i^2.
\te
For this metric the Lagrangian density of a massless conformally coupled scalar
field is
\begin{equation}
{\cal
L}(x)=\frac{a^3}{2}\left[(\dot{\Phi})^2-\frac{1}{a^2}\sum_{i}(\Phi_{,i})^2
-\left(\frac{\dot{a}^2}{a^2}-\frac{\ddot{a}}{a}\right)\Phi^2\right]
\end{equation}
where a dot denotes a derivative with respect to $t$.
Decomposing $\Phi$ in normal modes we find (after adding a surface term)
\begin{equation}
L (t)=\sum_{\sigma}^{+-}\sum_{\vec{k}}\frac{a^3}{2}
\left[(\dot{q}_{\vec{k}}^{\sigma})^2+2\frac{\dot{a}}{a}\dot{q}_{\vec{k}}^{\sigma}q_{\vec{k}}^{\sigma}-\left(\frac{k^2}{a^2}-\frac{\dot{a}^2}{a^2}\right) q_{\vec{k}}^{\sigma 2}\right]
\end{equation}
where $k$=$|\vec{k}|$ and $L(t)=\int{\cal L}(x)d^3 \vec{x}$.
If the detector- field interaction is of the same form as (5.6), but at fixed
$x$, we see that in a four-dimensional field the spectral density is given by
\be
I(k)=\left(\frac{\e}{2\pi}\right)^2 k.
\te
Using the Lagrangian (5.17) we find from (4.2) that the Bogolubov coefficients
are
\be
\alpha=\frac{(1+a^2)}{2a}e^{-ik\eta},\;\;\beta=\frac{(1-a^2)}{2a}e^{-ik\eta}
\te
where $\eta=\int_{t_i}^t dt/a(t)$ with $a(t_i)=1$. Using these we find that
the noise and dissipation kernels (4.9-10) are
\be
\zeta(t,t')=\nu(t,t')+i\mu(t,t')=\frac{1}{a(t)a(t')}\int_0^{\infty}dk
{}~I(k)e^{-ik(\eta-\eta')}.
\te
We will now specialise to the de Sitter space where, in the spatially-flat
RW coordinatization \cite{BirDav}, the scale factor has the form
\be
a(t)=e^{Ht}.
\te
In this case $\eta=-\frac{1}{H}e^{-Ht}$ with $t_i=0$. If we define
$\Delta=t-t',\;\Sigma=t+t'$ we find that (5.20) becomes
\be
\zeta(t,t')=e^{-H\Sigma}\int_0^{\infty}dk ~I(k)
\exp\left[-\frac{2ik}{H}e^{-H\Sigma/2}\sinh (H\Delta/2)\right].
\te
Making use of \cite{Ang}
\be
e^{-i\alpha\sinh (x/2)}=\frac{4}{\pi}\int_0^{\infty}d\nu
K_{2i\nu}(\alpha)[\cosh (\pi\nu)\cos (\nu x)-i\sinh (\pi\nu)\sin (\nu x)]
\te
we find that
\be
\zeta(t,t')=\int_0^{\infty}dk~G(k)\left[\coth \left(\frac{\pi k}{H}\right)\cos
k(t-t')-i\sin k(t-t')\right]
\te
where
\bea
G(k)&=&\frac{4\sinh (\pi k/H)}{\pi H
e^{H\Sigma}}\int_0^{\infty}dk'~I(k')K_{2ik/H}(2k'e^{-H\Sigma/2}/H) \nn \\
&=&\left(\frac{\e}{2\pi}\right)^2 k=I(k).
\tea
We have used the integral identity
\be
\int_0^{\infty}dx~x^{\mu}K_{\nu}(ax)=2^{\nu-1}a^{-\mu-1}\Gamma\left(\frac{1+\mu+\nu}{2}\right)\Gamma\left(\frac{1+\mu-\nu}{2}\right)
\te
and the properties of gamma functions.
Comparing (5.24) with (5.4) we see that
a thermal spectrum is detected by an inertial observer in de Sitter space at
temperature
\be
k_B T= {H \over {2 \pi}}.
\te

\section{Field-Spacetime Coupling: backreaction in semiclassical cosmology}

The QBM paradigm can be applied effectively to treat problems in semiclassical
gravity, i.e., quantum matter fields in a classical spacetime and its
backreaction on the dynamics of spacetime.
A well-studied example is the dissipation of anisotropy due to particle
creation in the early universe \cite{CalHu87}.
(For a description of how a study of this problem led to the discovery of
the relevance of the closed-time-path effective action method \cite{ctp}
and the influence functional \cite{FeyVer} formalism, see e.g.,
\cite{HuWaseda}.)
Here spacetime is regarded as the system and the quantum field
as the environment. The effect of the quantum field on the dynamics of
spacetime is captured in the influence functional in the same way as
the oscillator bath on the Brownian particle. The motivation is explained
in \cite{HPZ1,HPZ2,HuBelgium}
and the details can be found in \cite{CalHuSG,HM3}.

Consider now another model with the action
\be
S[a,q]
 = \int\limits_{t_i}^tds \Biggl[ L_g(a,\dot{a},s)
 + \sum_n\Bigl\{  {1\over 2}m_n(a,\dot{a})
  \Bigl(\dot q_n^2+b_n(a,\dot{a})q_n\dot{q}_n-
\omega^2_n(a,\dot{a})q_n^2\Bigr) \Biggr]
\te
where $L_g$ is the classical gravitational action.
This action can be used to describe a free quantized scalar field propagating
in a spatially flat Friedmann-Robertson-Walker (FRW) universe with
scale factor $a(s)$. The field and spacetime are
coupled parametrically: The time dependence of the mass and frequencies of the
oscillators comes from the scale factor (system  variable) whose dynamics
is determined by the Einstein equations. The detector is left out of the
picture, replacing it  formally is the scale factor of the universe.
By tracing out the scalar field
we can obtain an influence functional from which an equation of motion
for the scale factor in the semiclassical regime can be derived.
We find that for an initial vacuum state the exact influence functional
for this model is \cite{HM3,CalHuSG}

\begin{equation}
{\cal F}[a,a']=\prod_n\frac{1}{\sqrt{\alpha_n
[a']\alpha_n^*[a]-\beta_n[a']\beta_n^*[a]}}.
\end{equation}
The Bogolubov coefficients are derived as before via equations (4.2),
where now $g$ and $h$ are functions of the system variable $a$.
The Bogolubov coefficients describe particle creation in each normal mode
$n$ of the scalar field generated by the expansion of the universe.

The influence functional
tells us how  particle creation reacts back on the dynamics of the universe
and determines the time-dependence of the scale factor $a$.
Using the influence functional method we have studied the backreaction in
detail
\cite{CalHuSG,HM3} and
obtained in the semiclassical limit an equation of motion for $a$ which
includes non-local dissipation and colored noise. The vacuum fluctuations of
the scalar field is the source of the noise.
This approach to semiclassical gravity goes beyond the conventional
semiclassical Einstein equation, which
considers only the average value of the matter energy momentum tensor.
By approaching semiclassical gravity from the statistical mechanical viewpoint
we see how all statistical information in the quantum matter source,
in particular the noise and fluctuations of quantum fields,
can be systematically included in the backreaction. Recognition of the
stochastic nature of semiclassical gravity is important because it
represents a qualitative change in the theory which can
lead to a deeper understanding of all semiclassical gravity phenomena.

\section{Discussion}

This talk aims at addressing the origin and nature of  noise in
quantum systems. The methodology and concepts we introduced can be
used to tackle a wide range of problems, such as those listed in the
Introduction.

On a practical level there are at least two simple advantages
in discussing noise in gravitation and cosmology:\\
1) The frequent occurance and ubiquitous nature of colored noise in cosmology
enrich the reportoire and enhance the function of noise in physical systems
as we have learned in this meeting.\\
2) The theory of noise developed in ordinary physical systems can be deepened
to probe into the statistical properties of the vacuum in curved spacetime.
 From this one can perform more in-depth analysis of the quantum statistical
processes in the early universe and black holes.

Although we have used examples from quantum cosmology to illustrate
the physical relevance of the quantum Brownian model with a
parametric oscillator bath, the
range of applicability of this model goes beyond. An important area
where parametric amplification in coupled oscillators plays a central
role is in quantum optics  \cite{Milburn}.
There it is believed that squeezed baths can serve to retain certain quantum
behaviour of the system
in its interaction with the environment. The model studied in this work is
exact and its general results will therefore prove useful for addressing
similar issues in this context.

On a theoretical level,
the theory of noise and fluctuations lie at the foundations of statistical
mechanics and quantum field theory.
Some basic issues in gravitation and cosmology require the understanding
of both of these aspects. These two branches of theoretical physics
are however almost entirely disjoint until the discovery of Hawking effect
for black holes \cite{Haw75} and Unruh's discovery of thermal radiance in a
uniformly
accelerated observers \cite{Unr}.
Sciama \cite{Sciama}, in celebration of Einstein's other major accomplishment,
was the first to introduce a fluctuation theory viewpoint in
explaining these effects. This very insightful viewpoint has unfortunately
received little attention. From the work of one of us over the years
in trying to understand the nature of dissipation in semiclassical
gravity \cite{HuPhysica} we are led to focussing more on the effect of
noise and fluctuations in quantum fields as the source of decoherence
in the emergence of classical spacetime and dissipation in the spacetime
dynamics
\cite{HuTsukuba}.
Noise and fluctuation could also be instrumental in inducing a
phase transition at the Planck scale
between the phases described by a theory of quantum
gravity and the classical theory of general relativity \cite{HuBanff}.
We think this viewpoint can stimulate new ideas and open new channels of
fruitful investigations in gravitation and cosmology research.

 From this discussion you can perhaps understand why we said in the beginning
of the talk that noise in the cosmos is essential. From it particles are
created,
galaxies are formed and entropy is generated, not to mention their role in
`creating' the quantum universe and bringing about the comfortable
classical world we live in.
Actually considering noise as good or bad is rather irrevelant in this
regard, as we owe our existence to it in the final analysis.
Paraphrasing Voltaire: they simply exist and thrive in spite of us.

{\bf Acknowledgement} We thank Dr. Marko Millonas and other
organizers of this workshop for
a very interesting meeting. BLH would like to thank his
collaborators in this research program, Esteban Calzetta, Juan-Pablo Paz,
Sukanya Sinha and Yuhong Zhang, for many years of  stimulating
exchanges and rewarding collaboration. This research is supported in
part by the National Science Foundation under grant PHY91-19726.


\end{document}